\documentclass[runningheads,a4paper]{llncs}[2015/06/24]

\usepackage{cmap}
\usepackage[T1]{fontenc}

\usepackage{graphicx}

\usepackage[ngerman,english]{babel}
\addto\extrasenglish{\languageshorthands{ngerman}\useshorthands{"}}

\usepackage[%
rm={oldstyle=false,proportional=true},%
sf={oldstyle=false,proportional=true},%
tt={oldstyle=false,proportional=true,variable=true},%
qt=false%
]{cfr-lm}
\usepackage[math]{blindtext}

\usepackage{cite}

\usepackage{paralist}

\usepackage{csquotes}

\usepackage{amsmath}
\usepackage{amsfonts}

\usepackage{microtype}

\usepackage{url}
\makeatletter
\g@addto@macro{\UrlBreaks}{\UrlOrds}
\makeatother

\usepackage{xcolor}

\usepackage{pdfcomment}
\hypersetup{hidelinks,
   colorlinks=true,
   allcolors=black,
   pdfstartview=Fit,
   breaklinks=true}
\usepackage[all]{hypcap}

\newcommand\todo[1]{}
\newcommand\TODO[1]{}

\usepackage[capitalise,nameinlink]{cleveref}
\crefname{section}{Sect.}{Sect.}
\Crefname{section}{Section}{Sections}

\usepackage{booktabs}
\usepackage{braket}

\usepackage{listings}
\usepackage{why3lang}
\usepackage{lstautogobble}

\begin{document}

\newcommand{\why}{\emph{Why3}}
\newcommand{\coq}{\emph{Coq}}
\newcommand*{\rom}[1]{\uppercase\expandafter{\romannumeral #1\relax}}

\title{Experience Report: \\ Formal Methods in Material Science}
\author{Bernhard Beckert \and
Britta Nestler \and
Moritz Kiefer \and \\
Michael Selzer \and
Mattias Ulbrich}

\institute{Karlsruhe Institute of Technology}

\authorrunning{Beckert, Nestler, Kiefer, Selzer, Ulbrich}

\maketitle

\begin{abstract}
  Increased demands in the field of scientific computation require
  a more efficient implementation of algorithms. Maintaining
  correctness, in addition to efficiency, is a challenge software
  engineers in the field have to face.
  In this report, we share our first impressions and experiences with the
  applicability of formal methods to such design challenges arising in
  the development of scientific computation software, in the field of
  material science.
  We investigated two different algorithms, one for load distribution
  and one for the computation of convex hulls, and demonstrate how
  formal methods have been used to discover counterexamples to the
  correctness of the existing implementations, and to prove the
  correctness of a revised algorithm.
  The techniques employed include SMT solvers, and automatic
  and interactive verification tools.
\end{abstract}

\section{Introduction}

The demands on precision and the extent of simulations, and other
scientific computation applications increase continuously.
In contrast to these requirements, Moore's law for technological
advances of processors faces a foreseeable stagnation, which will make
a future development of more efficient software unavoidable and
necessary in this area.
%
With the requirement to design scientific software with more sophisticated
algorithms, using distributed computing, efficient memory
technologies, \textit{etc.}, the number of pitfalls grows, by which
flaws could accidentally be introduced into the code.

Bugs in scientific software are hard to discover by testing, since they
may only occur at inputs arising after a longer (simulation) runtime.
This challenge in software development thus appears to be an
opportunity for static formal analyses which analyse the code
symbolically, and thus cover all possible runtime situations --
regardless of the time required to reach the value during an actual
run.

%
In order to verify the hypothesis "Formal methods can be employed successfully
in the design process for scientific software", we tested it against
two case studies from real scientific applications, for which the
location of problematic code had previously been identified.
In one case, the challenge was to find a correct implementation which
meets the intention of the original code, without repeating its flaws.
In the other case, the flaw had already been corrected. The
challenge was to verify that the correction solves the existing
problem.

The paper continues with a presentation of the two case studies, which
is followed by a summary of the observations made during the analysis
process, and it concludes with an outlook into future work.

\section{Case Study \rom{1}: Load Distribution}
\label{sec:load-distr}
For an efficient use of resources, provided by large computing
clusters, load distribution is a crucial challenge.
In this case
study, we analyse an algorithm to rebalance the load within a cluster,
in case the number of tasks to be computed changes. An original version of
the algorithm in question has been in use in a scientific computation
software, and its correctness could not be established by manual code
inspection.

This document reports how formal analyses helped us to formulate a
correct load balancing algorithm by
\begin{enumerate}
\item finding a subtle flaw in original implementation, using a custom fuzzer, and by
\item proving the correctness of an improved implementation, using \why{}~\cite{why3}.
\end{enumerate}

\paragraph{Problem Statement}%
\label{sec:prob-stm-i}

The starting point for this case study was an algorithm taken from a
scientific computation library which was actively in use. The algorithm computes
the distribution of $\mathbf{s}$ tasks onto $\mathbf{n}$ cluster nodes
from a given distribution for $\mathbf{t}$ tasks. This routine is
called within a simulation framework when the number of simulated
entities is refined (usually increased).

More formally, the problem can be stated as follows:
Given a set $S=\{s_{1},\ldots,s_{n}\}$ with $s_{i}\in \mathbb{N}_{+}$
and $\mathbf{s} = \sum_{i=1}^{n} s_{i}$ and a natural number
$\mathbf{t}\in\mathbb{N}_{+}$, produce a set
$T=\{t_{1},\ldots,t_{n}\}$ such that
$\mathbf{t} = \sum_{i=1}^{n} t_{i}$.
The respective values $s_i$ and $t_i$ are the number of tasks to be run on the
$i$-th node, before and after load balancing.
The objective of the algorithm is to obtain a new load distribution that is
``close'' to the old distribution, i.e., $\frac{t_{i}}{s_{i}}$ should
be close to $\frac{\mathbf{t}}{\mathbf{s}}$.

\paragraph{Starting Point}

The original algorithm shown in Listing.~\ref{lst:original} used
floating-point numbers to calculate
$s_{i}\frac{\mathbf{t}}{\mathbf{s}}$ for each $i$. The integral part
of this value then was assigned to $t_{i}$, while the fractional parts
were accumulated until they made up $1$ full node, which was then
assigned to the current node.
This algorithm computes a correct distribution among the nodes, if
executed on precise rational numbers. However, when using floating-point
values to approximate rationals, the imprecisions can 
result in tasks getting lost in the balancing process, i.e., it can be
that $\sum_{i=1}^{n} t_{i} 
< \mathbf{t}$. \todo{Mention
  that it is also possible (is it?) for this value to be larger}

\paragraph{Automatic discovery of counterexamples}

\begin{figure}[t]
    \centering
  \begin{tabular}{rrrl}
    \toprule
    \(s_{1}\) & \(s_{2}\) & \(\mathbf{t}\) & rest \\
    \midrule
    1048627  &    524206 & 1099511627744 & 0.9998779296875 \\
      32779  & 536870892 & 1099511627779 & 0.999881774187088 \\
    67108824 &  33554439 & 1099511627792 & 0.9998779296875 \\
    \bottomrule
  \end{tabular}
  \caption{Counterexamples to the correctness of the original algorithm.}\label{fig:counterexamples}
\end{figure}

Providing concrete counterexamples first requires a definition of the
\lstinline|isNearlyEqual| method. In the original implementation, this
function was implemented as an absolute \(\epsilon\)-comparison using
\lstinline|FLT_EPSILON| as the value for \(\epsilon\). \lstinline|FLT_EPSILON|
is defined as the difference between 1 and the smallest floating-point
number of type \lstinline|float| that is greater than
1~\cite{glibc-fp}. Under the assumption that \lstinline|float| refers to the 32bit
floating-point type defined in IEEE754, as is usually the case,
this value is \(2^{-23}\).

A careful analysis leads to the conclusion that it should be
possible to find a counterexample with \(n=2\). However, the search
space consisting of the three \lstinline|long| values \(s_{1}\),
\(s_{2}\) and \(\mathbb{t}\) is still too large to be explored
exhaustively. Thus, we first reduced the search space further, and then
used random fuzzing to discover counterexamples. To reduce the search
space, we drew the values \(s_{1},s_{2},\mathbf{t}\) randomly from
\(\Set{ 2^{e} + \delta | e \in \Set{0,\ldots,40}, \delta \in
  \Set{-100,\ldots,100}}\). The rationale for this choice is that
small offsets from large integer powers use a large range of the precision,
and thereby are likely to lead to imprecisions during calculations.

The fuzzer finds numerous counterexamples within a matter of
seconds, a few of which can be seen in Fig.~\ref{fig:counterexamples},
including the final value of \lstinline|rest| which is below
\(0.9999\), and thereby is well below \lstinline|1 - FLT_EPSILON|.

\paragraph{Improved Algorithm}

To remedy the problems in the original algorithm, we eliminated all
uses of floating-point numbers in the original algorithm in favour of
integer operations. Then we verified three different
properties of this algorithm using \why{}:

\begin{enumerate}
\item No tasks are lost, i.e., $\sum_{i=1}^{n} t_{i} = \mathbf{t}$.
\item The resulting values $t_{i}$ are close to
  $s_{i} \frac{\mathbf{t}}{\mathbf{s}}$. In particular, the following
  holds: \\
  $\lfloor \frac{\mathbf{t}}{\mathbf{s}} \rfloor \leq
  \frac{t_{i}}{s_{i}} \leq \lceil \frac{\mathbf{t}}{\mathbf{s}}
  \rceil$.
\item The integer algorithm is equivalent to the original algorithm, if
  rationals are used instead of floating-point values (i.e., if all
  computations are exact, no rounding effects).
\end{enumerate}

Property 3 is particularly interesting, since proving the functional
equivalence of the original algorithm to the new integer algorithm
ensures that the intentions of the original author, which can include
domain knowledge not available to the verification engineer, are
preserved.

While the verification of property 1 was possible automatically using
\why{} after adding lemmas to assist the proof search, property 2 and
3 required reasoning about properties, involving 
floating-points, which is a weakness of most automatic theorem
provers. We thus had to resort to time-intensive interactive proofs using
the \coq{} theorem prover~\cite{coq}.

\begin{figure}[t]
\begin{minipage}[t]{0.49\textwidth}
\begin{lstlisting}[language=C,captionpos=b,caption=Original algorithm,label={lst:original}]
  double rest = 0.0;
  for (int i = 0; i < num_tasks; ++i) {
    double share =
      (double)tasks[i]/(double)total_tasks;
    double real_size =
      share * (double)new_total_tasks;
    double floor_size = floor(real_size);
    rest += real_size - floor_size;
    new_tasks[i] = (int)floor_size;
    if (isNearlyEqual(rest, 1.0)) {
      new_tasks[i] += 1;
      rest -= 1;
    }
  }
\end{lstlisting}
\end{minipage}
\begin{minipage}[t]{0.49\textwidth}
\begin{lstlisting}[language=C,captionpos=b,caption=Revised algorithm,label={lst:revised}]
  int rest = 0;
  for (int i = 0; i < num_tasks; ++i) {
    int scaled = new_total_tasks * tasks[i];



    int floor_size = scaled / total_tasks;
    rest += scaled % total_tasks;
    new_tasks[i] = floor_size;
    if (rest >= total_tasks) {
      new_tasks[i] += 1;
      rest -= total_tasks;
    }
  }
\end{lstlisting}
\end{minipage}

\caption{Original algorithm using floating-points and the revised
  algorithm which has been verified. The input to both programs are
  two pointers \lstinline|tasks| and \lstinline|new_tasks|, the number
  of tasks \lstinline|num_tasks| and the total number of tasks in the
  \lstinline|tasks| array \lstinline|total_tasks| as well as the
  number of tasks \lstinline|new_total_tasks| which should be in
  \lstinline|new_tasks| at the end of the algorithm.}
\end{figure}

\begin{lstlisting}[float,floatplacement=b,language=why3, caption={The
  specification and implementation of the \emph{resize} algorithm in
  Why3. Note that this only specifies properties~1 and~2. For
  presentational purposes we have omitted imports from the standard
  library of \why{}, and reuse the same arithmetic operators for integers and
  reals. \why{} can verify property 1~automatically using the supplied
  lemmas while property 2~requires interactive proofs in the theorem prover \coq{}.}]
  module Resize
    lemma small_rest : forall n:int, a:int, b:int, c:int.
      n>=0 /\ a>=0 /\ b>=0 /\ c>=0 /\ c<n /\ n*a=n*b+c -> c=0
    lemma floor_ceil : forall r:real.
      from_int (floor r) < r -> floor r + 1 = ceil r
    lemma floor_rest : forall i:int, r:real.
      i>=0 /\ 0.0<=r /\ r<1.0 -> floor (from_int i + r) = i
    lemma floor_div : forall a:int, b:int.
      a>=0 /\ b>0 -> div a b = floor (from_int a / from_int b)
    lemma floor_le_ceil : forall r:real. floor r <= ceil r

    let resize (tasks : array int) (total_tasks: int) (new_total_tasks: int)
      requires { total_tasks = sum tasks 0 (length tasks) }
      requires { forall i:int. 0 <= i < length tasks -> tasks[i] >= 0 }
      requires { 0 < total_tasks /\ 0 <= new_total_tasks }
      ensures { new_total_tasks = sum result 0 (length result) }
      ensures {
        forall i:int. 0 <= i < length result ->
          let exact = from_int tasks[i] * from_int new_total_tasks /
                      from_int total_tasks
          in floor exact <= result[i] <= ceil exact
      }
    = let new_tasks = make (length tasks) 0 in
      let rest = ref 0 in
      for i = 0 to length tasks - 1 do
        invariant { sum tasks     0 i * new_total_tasks
                  = sum new_tasks 0 i * total_tasks + !rest }
        invariant { sum new_tasks 0 i >= 0 }
        invariant { 0 <= !rest < total_tasks }
        invariant { forall j:int. 0 <= j < i ->
            let exact = from_int tasks[j] * from_int new_total_tasks /
                        from_int total_tasks
            in floor exact <= new_tasks[j] <= ceil exact
        }
        let floor_size = div (new_total_tasks * tasks[i]) total_tasks in
        rest := !rest + mod (new_total_tasks * tasks[i]) total_tasks;
        new_tasks[i] <- floor_size;
        if !rest >= total_tasks then (
          new_tasks[i] <- new_tasks[i] + 1;
          rest := !rest - total_tasks
        );
      done;
      assert { total_tasks * new_total_tasks =
               total_tasks * sum new_tasks 0 (length tasks) + !rest };
      assert { !rest = 0 };
      new_tasks

  end
\end{lstlisting}

\section{Case Study \rom{2}: Convex Hull}

Calculating the convex hull of a set of points is a problem which is
commonly found in scientific computation applications.
However, while the algorithms for the computation of the hull are mathematically
simple and straightforward, numerical errors caused by using floating-point
values in implementations instead of real numbers, can lead
to a wrong result.

In the two-dimensional case, the implications are not too severe: It
might be that a point close to an edge of the convex hull is wrongly
included in (or wrongly excluded from) the hull polygon. But the result is
always a valid polygon which is close to the desired result.

The situation is different if three-dimensional data is taken as
input. The additional dimension requires keeping a record of
the facets making up the convex polyhedron. This situation is computationally
considerably more sphisticated than the two-dimensional case 
as it requires the calculation of the side of a facet (front or back) that is faced by a point. 
For points close to the facet, such a calculation may
come up with the wrong result due to floating-point imprecision.
The situation becomes bad if this computation for two facets errs for one point in
different directions: Then the convex hull, which relies on
these computations, wrongly includes or excludes facets from the hull,
with the catastrophic outcome that the result is not only imprecise,
but not a (closed) polyhedron at all -- which is a far more severe
problem.

This problem, which was reported by Barber~et.~al.~\cite{quickhull}, when presenting Quickhull, a widely employed algorithm for convex hull computation, has been known for a long time.


The existing implementation given here was known to
suffer from such errors, and it was also known that errors had occurred in practice.
As workarounds were implemented into the code to mitigate the problem,
the number of observed errors decreased, but the effectiveness of the
solution for the general case was unclear.
The workarounds are focused on a particular method which given a plane
spanned by three points and a fourth point decides whether that point
is above, below or coplanar to the plane.

As mentioned above, due to floating-point imprecision effects, the
wrong decision might be made with fatal effects.
In an attempt to avoid faulty results due to rounding errors, the
original method has been modified to compute the output three times.
Each time, a different point is chosen from the three spanning points
as the base point of the plane.
%
As final result, the result computed by the majority, is returned.

In this case study, we did not apply formal methods to prove a
correctness hypothesis, like in Section~\ref{sec:load-distr}, but 
we succeeded in \emph{disproving} that the
introduced workaround works in all cases.
We used the state-of-the art SMT solver \emph{Z3}~\cite{z3} to find
inputs were the majority vote computes a different result from the
exact result (on real numbers, not floats).
Since this verification task heavily depended on the semantics of the
floating-point arithmetic, which is very difficult to handle in a
deductive fashion, we chose to model it using the SMT solver. This
technology (in almost all cases) reduces problems on floating-point
values to problems on corresponding bitvectors according to the
standard IEEE~754. These are then resolved into an instance of the
propositional SAT problem which can then be solved using a Boolean
satisfiability solver.

While the search was successful, it required a careful
analysis of the problem to reduce the search space.
Floating-point problems are known to have the tendency to lead to
SAT instances that are difficult to solve. A thorough manual analysis of the situation
allowed us to narrow the search space for the floating-point values
considerably.
Even with the reduced search space, \emph{Z3} required
approx.~100\,h\footnote{on a virtualised Intel Xeon E3 with 2.6\,GHz}
to find a counterexample.
On the other hand, was virtually impossible to manually come up with
concrete numbers that constitute a counterexample, because the
dependencies of individual bits in floating-point arithmetic are
difficult to follow.

\section{Observations}

In this report, we have looked at how formal methods can be
applied to two orthogonal problems. The formal techniques 
employed in this study can be separated into finding counterexamples to
disprove the fact that existing algorithms satisfy certain properties, and into
verifying the correctness of algorithms. The latter can again be separated
into using formal verification to prove the correctness, as given by an
abstract specification, and into relational techniques which prove the
equivalence to an existing algorithm.

While it is still too early to draw definitive conclusions from these
early investigations, they give rise to some interesting observations
of the potential applicability of formal methods in the field.

One interesting observation is that it is often possible to isolate
the problems faced in large projects to relatively small and isolated
pieces of code. While the formal analysis of the original project
might not be feasible due to size and complexity, analysing these
small pieces of code is significantly easier. However, extracting such
a piece of code is a process that often requires domain
knowledge. The provision of tools which help or automate this extraction, such
that it can be done by the domain experts themselves, could reduce the
time needed to produce code samples suitable for formal analysis.

While we have been able to find counterexamples automatically using
SMT solvers, this relies heavily on an upfront reduction of the search
space, which requires experience with formal methods, and cannot be
applied by scientists themselves. Automating this process such that it
could be applied by the developers of the original algorithm would be
highly desirable, but requires further research. When comparing the use of
fuzzers and SMT solvers to find counterexamples, fuzzers are
significantly better at providing fast feedback which is suitable for
interactive use. For intricate problems, random
fuzzing is not likely to discover edge cases, on the other hand. 
A combination of both SMT solvers and fuzzers, while
only requiring a single specification, could help to combine the
respective benefits.

Overall, our experience shows that formal techniques can be employed
successfully to assist scientists. However, this relies on the ability
to isolate problems to smaller code samples, which is a tedious
process. Furthermore, while the large effort required for a formal
verification can be justified for algorithms used for longer
periods of time, immediate feedback provided for scientists during
development would be highly desirable, but is not easily achievable
using existing techniques.

\bibliographystyle{splncs03}
\bibliography{paper}

\begin{thebibliography}{1}
\providecommand{\url}[1]{\texttt{#1}}
\providecommand{\urlprefix}{URL }

\bibitem{quickhull}
Barber, C.B., Dobkin, D.P., Huhdanpaa, H.: The quickhull algorithm for convex
  hulls. ACM Trans. Math. Softw.  22(4),  469--483 (Dec 1996),
  \url{http://doi.acm.org/10.1145/235815.235821}

\bibitem{why3}
Bobot, F., Filli\^atre, J.C., March\'e, C., Paskevich, A.: Why3: Shepherd your
  herd of provers. In: Boogie 2011: First International Workshop on
  Intermediate Verification Languages. pp. 53--64. Wroc\l{}aw, Poland (August
  2011), \url{https://hal.inria.fr/hal-00790310}

\bibitem{z3}
de~Moura, L., Bj{\o}rner, N.: Z3: An efficient smt solver. In: Ramakrishnan,
  C.R., Rehof, J. (eds.) Tools and Algorithms for the Construction and Analysis
  of Systems. pp. 337--340. Springer Berlin Heidelberg, Berlin, Heidelberg
  (2008)

\bibitem{glibc-fp}
glibc~development team:
  \url{https://www.gnu.org/software/libc/manual/html_node/Floating-Point-Parameters.html},
  [Online; accessed 01-February-2017]

\bibitem{coq}
development team, T.C.: The Coq proof assistant reference manual. LogiCal
  Project (2004), \url{http://coq.inria.fr}, version 8.6

\end{thebibliography}

\end{document}